# Technological maturation of a Gigapixels Astrometry instrument for the Habitable World Observatory

Florence Ardellier-Desages (1), Jerôme Amiaux (1), Jérôme Bobin (5), Stéphane Chevobbe (5), Eric Doumayrou (1), Pierre-Antoine Frugier (1), Pierre-Olivier Lagage (1), Manon Lizzana (2,3), Fabien Malbet (2), Jerôme Martignac (1), Julien Michelot (4), Fabrice Pancher (2), Thibault Pichon (1), Samuel Ronayette (1), Hugo Rousset (2).

(1) Univ. Paris-Saclay, Univ. Paris Cite, CEA, CNRS, AIM, Gif sur Yvette, France

(2) Univ. Grenoble Alpes, CNRS, IPAG, Grenoble, France

(3) Sorbonne Universite, CNRS, Institut d'Astrophysique de Paris, Paris, France

(4) Pyxalis, 170 Rue de Chatagnon, 38430 Moirans, France

(5) CEA, Univ. Paris-Saclay, Irfu-DeDip, Gif-sur-Yvette, France

## ABSTRACT

This paper presents the development strategy for HWO-AGATE (Accurate Guiding and Astrometry Targeting Exo-earths), a high-precision relative astrometry instrument concept designed for the future Habitable Worlds Observatory (HWO). The instrument is based on a gigapixel-class focal plane that incorporates the new generation of large-format GigaPyx CMOS sensors developed by Pyxalis. The proposed architecture relies on four 220-megapixel sensors assembled into a focal plane array and coupled with a dedicated calibration system to achieve the astrometric precision required for exoplanet detection and characterization. Preliminary study identified three major technological challenges. First, the performance of the 220-MP detectors must be validated against HWO's astrometric requirements. Second, the development of space-qualified packaging and integration solutions is required for a multi-sensor focal plane, while addressing the electrical and thermal constraints associated with the fast readout of large-format CMOS detectors. Third, an onboard calibration data-processing architecture must be implemented to reduce the raw data volume by approximately two orders of magnitude, enabling efficient operation within spacecraft resource limits. To address these challenges, we propose a comprehensive technology maturation program based on the construction of a representative end-to-end detection chain. The demonstration setup will include a partial focal plane equipped with 220-MP sensors, front-end electronics, a dedicated processing unit, and the calibration subsystem. The objective is to demonstrate compliance with HWO-AGATE requirements and to raise the maturity of large-format CMOS detector technology to Technology Readiness Level 5 (TRL 5). Detector validation and space-qualified packaging activities are already supported through institutional funding and the French national PEPR ORIGINS program in the RETINA (Researching Exoplanets Through Imaging for Narrow-angle Astrometry) project.

## 1. INTRODUCTION

The Habitable Worlds Observatory (HWO) is at the forefront space-based astrophysics and exoplanetary science[1], [2]. One of its most ambitious goals is to directly image and spectroscopically characterize dozens of exoplanet Earth-mass exoplanets candidates in their habitable zone (Exo-Earths). To characterize Earth like exoplanets, we propose developing an innovative relative astrometry instrument called HWO-AGATE (*Accurate Guiding and Astrometry Targeting Exo-earths*). This instrument will use large CMOS sensors to measure the mass of an Earth-mass exoplanet orbiting a Sun-like star with about 10% precision. The instrument concept relies on required parameters, state-of-the-art technology and environmental constraints.

## 2. SCIENCE DRIVERS AND SYSTEM DESCRIPTION

Relative astrometry consists in tracking the angular displacement of the target star's photocenter around the system barycenter relative to reference calibration stars. The primary science driver for a relative astrometric instrument is the detection of exo-Earth signatures at 10 pc[3]; the required precision is assessed at 0.3 µas level. Considering a target Signal-to-Noise Ratio (S/N) of 6, the end-of-mission astrometric noise floor is evaluated at 0.05 µas. Assuming an astrometric survey consisting of 100 visits per target, the single-measurement precision requirement is established at 0.5 µas. Astrometric instrument key parameters have been assessed to measure the offset angles between the target object at the center and the reference calibration stars[4].

Main values considered are summarized below:
- Visible instrument: 420 – 940 nm
- Small pixels : size below 4.5 µm to sample well the PSF
- Field of View of 12 arcmin$^2$
- High dynamic (up to 90 dB) and fast readout (up to 150 fps)
- Operation temperature : 290 K

Based on these parameters, an initial layout has been developed; it is based on a new, large CMOS matrix with a 4.4 µm pixel pitch, produced by the Pyxalis company, as part of the GigaPyx family. HWO-AGATE is designed around an Integrated Focal Plane included its front-end readout, a Data Processing Unit, a Power Control Unit, a Memory Unit and a calibration unit to achieve HWO astrometry requirements.
HWO-AGATE instrument is made of three main subsystems:

(1) The HWO – AGATE instrument is located at the telescope focal plane and covers a field of view of 3x4 arcmin$^2$. Its architecture is based on 4 large-format CMOS detectors and 3 smaller, faster CMOS detectors in the center (see Figure *1*). Specifically, Pyxalis provides 220 MP (megapixel) arrays with a 4.4 µm pixel pitch. Integrating these with 3 units of 14 MP smaller-scale arrays within a hybrid focal plane array (FPA) successfully addresses the demanding performance requirements of the HWO relative astrometry instrument. The front-end electronics are located as close as possible to the sensor area to minimize noise and to optimize signal integrity.

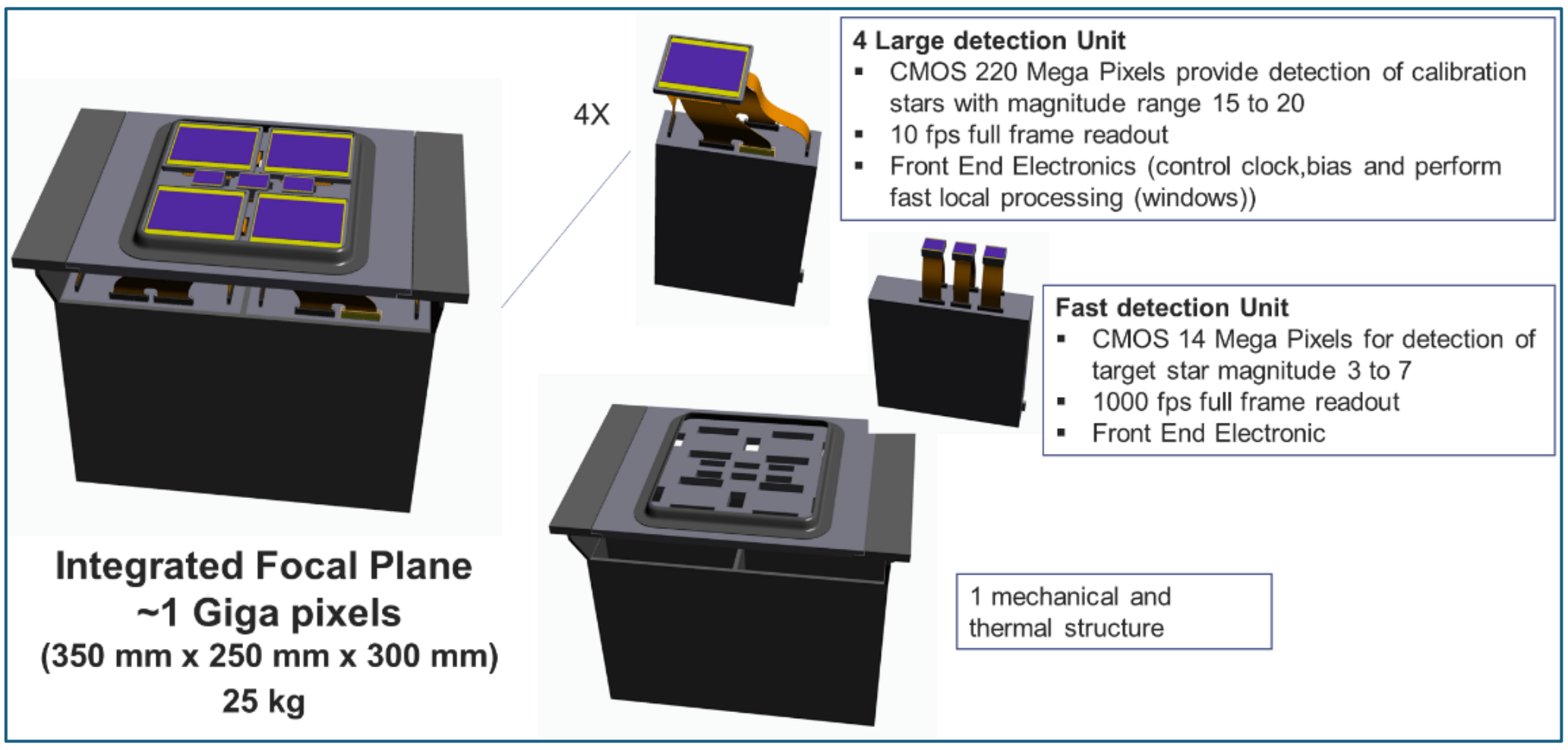

Figure 1 : HWO-AGATE Focal Plane architecture

(2) The HWO-AGATE Electronics Units composed of three independent units (see Figure *2*):
   a. Data Processing Unit (DPU) : process data splitted in science windowing and full frame calibration data reduction
   b. Power Control Unit (PCU): interface the Space Craft (S/C) primary power to provide secondary power to and Data Processing Unit (DPU).
   c. Memory Unit (MEU): Instrument local memory dedicated calibration processing

The final choice of a three-unit architecture is compatible with the mechanical and thermal architecture constraints of space missions. One of the major challenge for HWO-AGATE is on-board data management. The downlink data volumes sent to the ground segment a factor of 10 to 100 increase relative to current astrophysics missions such as EUCLID. Preliminary electrical architecture trade-off studies are underway to evaluate advanced packaging solutions, especially chiplet-based technologies, to meet the instrument's onboard processing requirements[4].

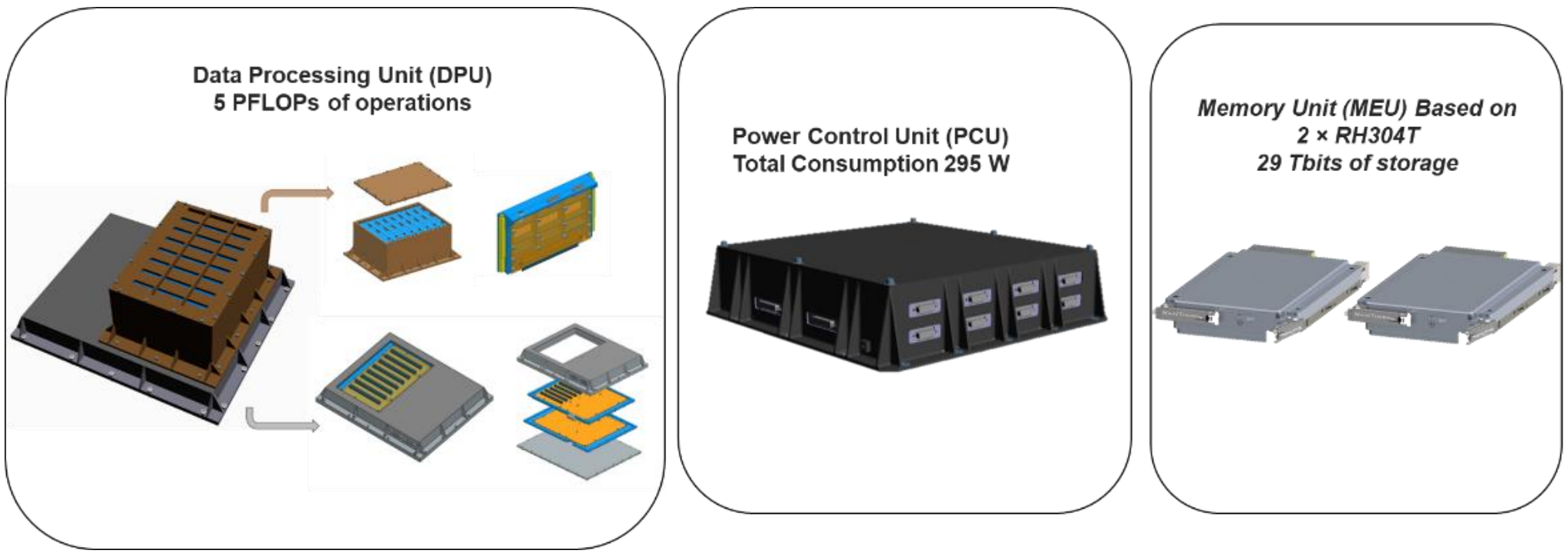

Figure 2 : Layout of HWO-AGATE Electronics and data management on board

(3) The HWO-AGATE Calibration System consists of a laser modulation unit and a fiber tip frame. It operates in the IPAG laboratory, and its global architecture must evolve to become space-qualified. (see Figure *3*).

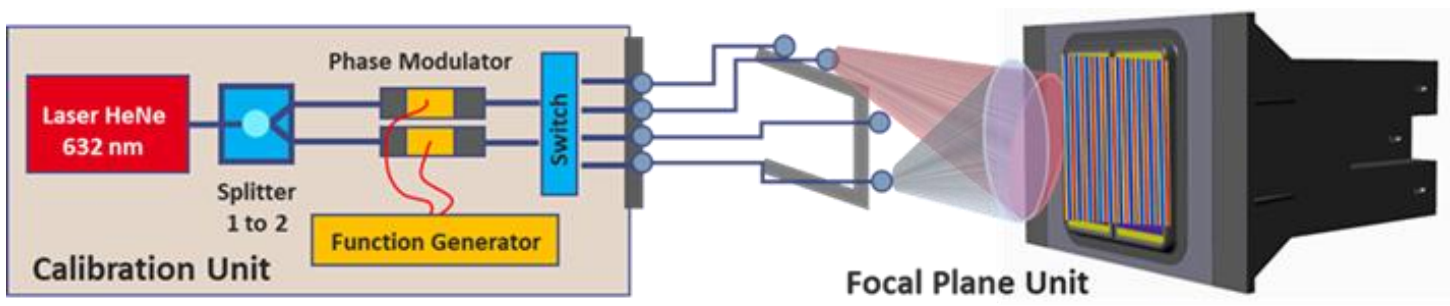

Figure 3 : Focal plane calibration layout

HWO-AGATE is probably one of the first concepts based on a gigapixel CMOS focal plane sensor designed for a space-based astrometry mission at the microarcsecond scale. Using large CMOS detectors represents breaks with traditional CCD architectures. The challenges lies not only in the detectors themselves, but also in their integrated management, packaging, calibration and on-board processing.

This paper describes the steps of the maturation plan to reach Technology Readiness Level (TRL 5) for the "220 MP Gigapyx Astro" development, as well as the integration of a partial focal plane and its calibration system in a dedicated test bench.

## 3. TECHNOLOGICAL MATURATION PLAN

To meet the performance science requirements for the HWO-AGATE program and to be able to be on board of space mission, three key elements have been identified to be compliant with this challenging space mission:

- 220 MP Gigapyx CMOS matrices to be compliant with space environment
- On-board data reduction

A maturation plan is being implemented to achieve NASA Technology Readiness Level 5 (Figure *4*) by the target date of HWO Mission Concept Review, scheduled for the end of the decade. The maturation roadmap involves transitioning from laboratory breadboards to environmental testing of demonstrator models in a relevant thermal-vacuum environment.

Regarding the 220 MP GigaPyx CMOS sensor, a four-phase maturation roadmap has been established to reach TRL 5 (Figure *5*).

- 1. Performance validation of the large format CMOS sensors of GigaPyx family
- 2. development of 'Astro' packaging tailored for space-borne focal planes
- 3. Detection chain performance validation
- 4. Validation CMOS 220 MP Gigapyx « Astro » in relevant environment.

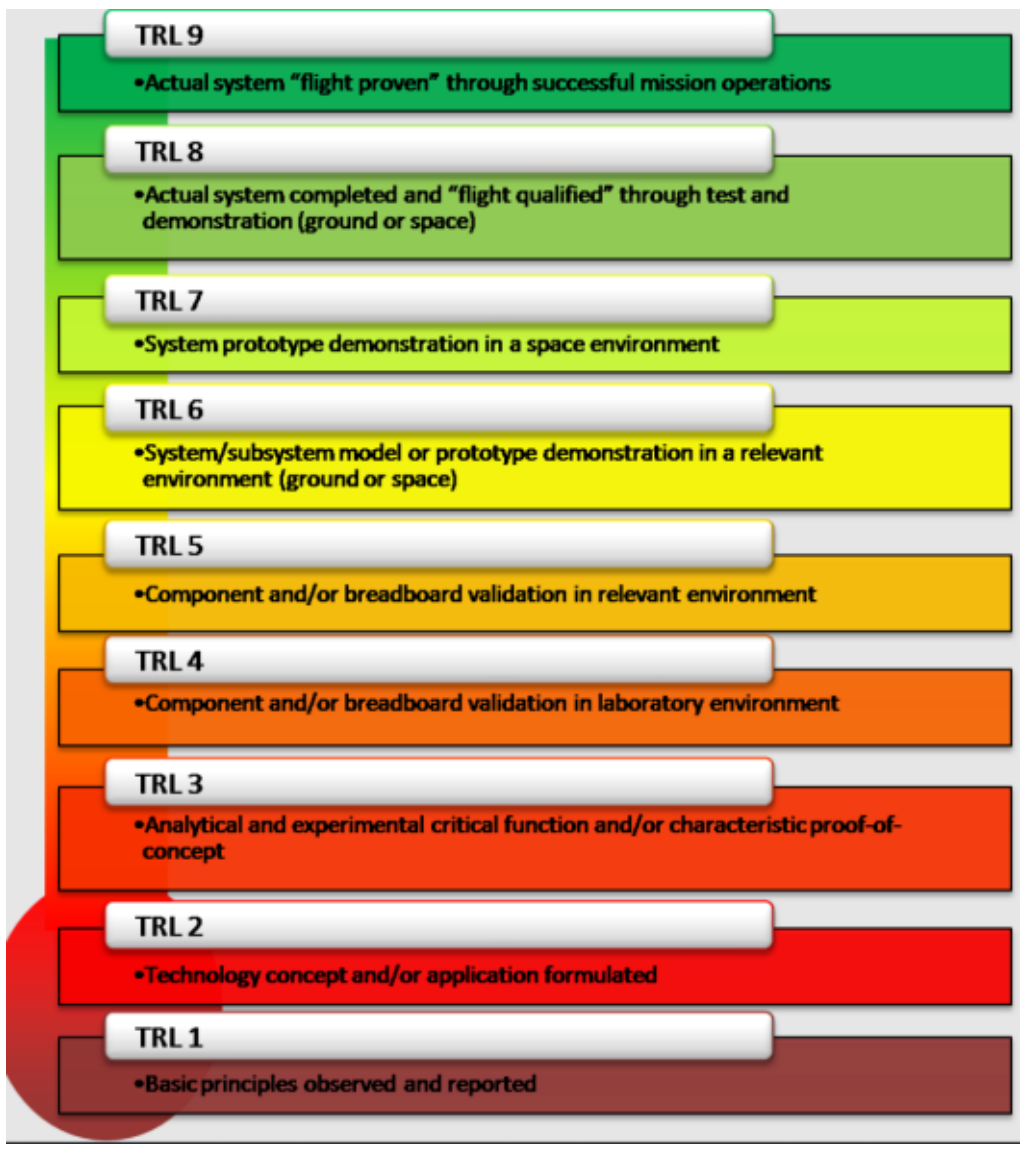

Figure 4: NASA Technology Readiness Level (TRL) scale

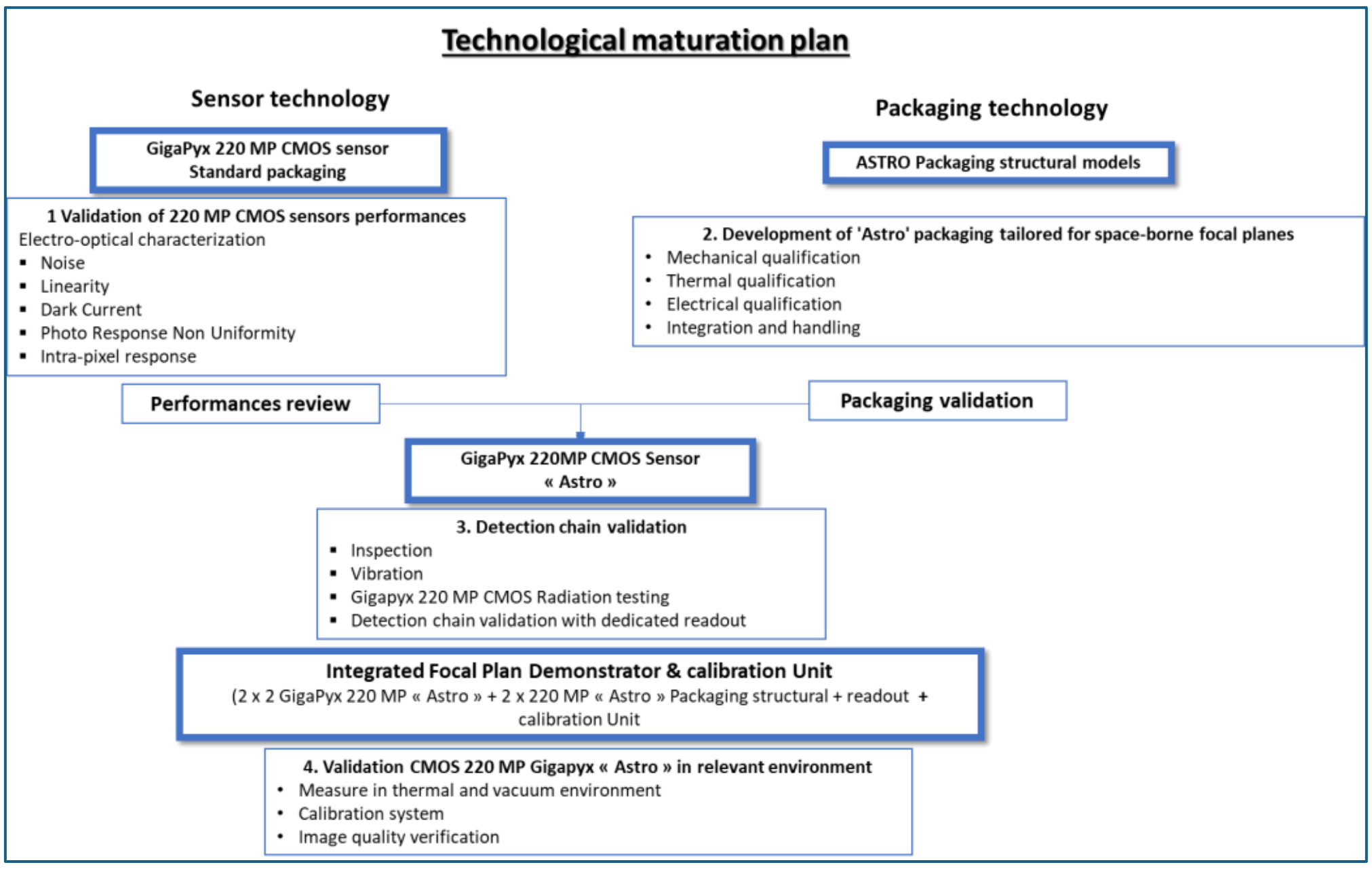

Figure 5: Technological maturation plan scheme

### 3.1 Phase 1: Validation of the 220 MP CMOS sensors performance

The objective of Phase 1 is to mitigate performances technological risk by demonstrating that the 220 MP CMOS sensor performances comply with HWO-AGATE requirements.
GigaPyx 220 MP CMOS matrices are fabricated by scaling GigaPyx 46 MP matrices using a photolithographic stitching process. A dedicated test bench developed at IPAG for the characterization of the GigaPyx 46 MP will be upgraded to test the 220 MP version. Performances measured so far on the 46 MP version are detailed in Lizzana et al. (2026)[5].

The first GigaPyx 220 MP sensors with standard packaging will be available first quarter 2027. They will be characterized using the same readout electronics and setup as the GigaPyx 46 MP to demonstrate that scaling the stitching process has not degraded the performances.

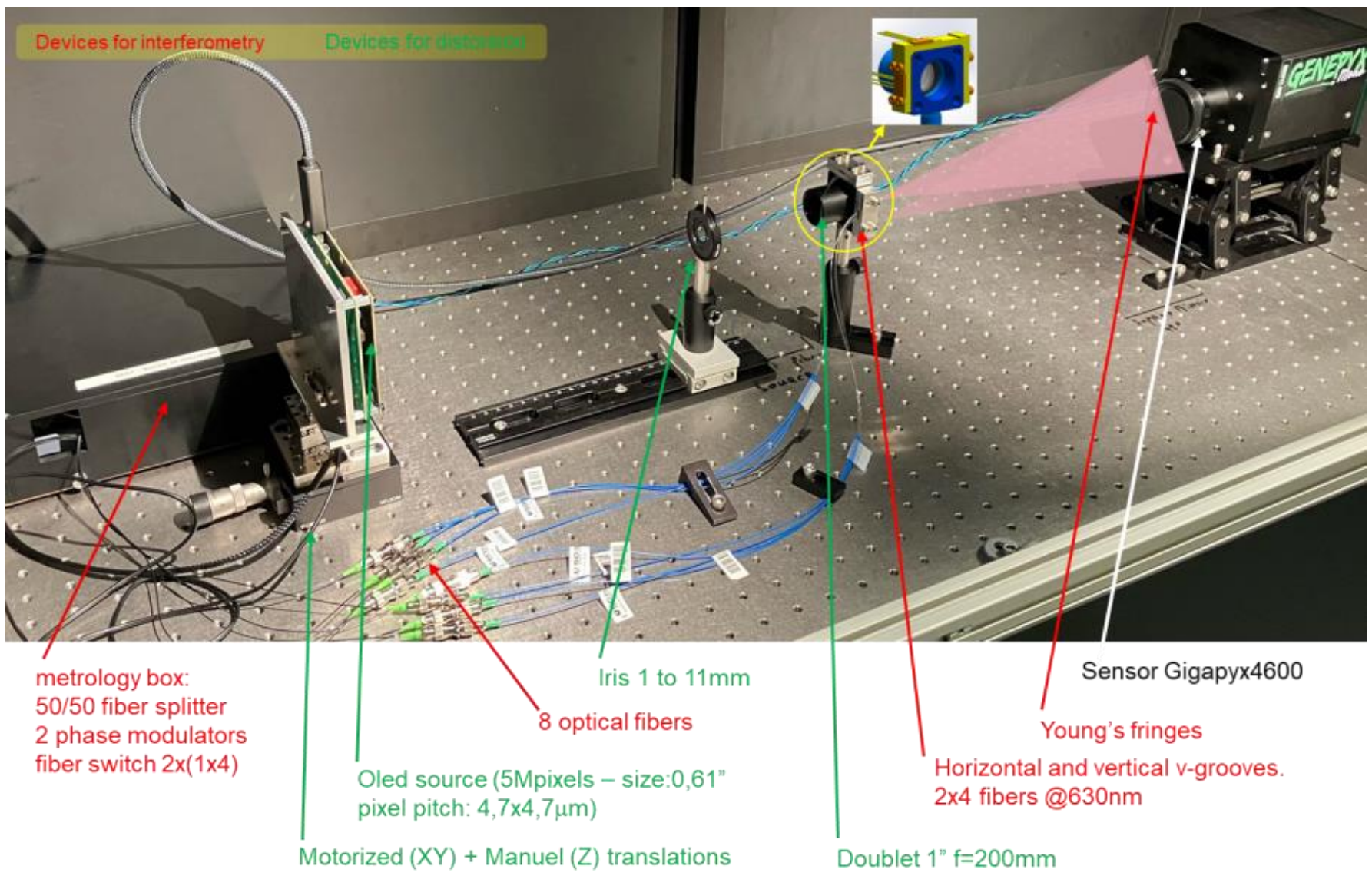

Figure 6: IPAG test bench

Figure *6* shows the IPAG tests bench that was used to characterize CMOS 46 MP Gigapyx detector[7], [8]. Table 1 summarizes the results.

*Table 1: GigaPyx 46 MP CMOS sensor performances (from Lizzana et al. 2026)* [5].

| Property | Pyxalis | IPAG |
|---|---|---|
| read out noise | 16 $e^-$ | 12 $e^-$ |
| linearity error | 2% | 0.90% |
| response non uniformity | $\leq$ 2% | 1.3% |
| dark current (at 35°C) | 24 $e^-$/s | 36 $e^-$/s |
| saturation charge | >50 k$e^-$ | 56 k$e^-$ |
| percentage of defective pixels | $\leq$ 0.1% | 0.0004% |
| inter-pixel capacitance | - | $\leq$ 3% |

Moreover, thanks to IntraPix tests bench, deeper intrapixels characterizations will be done. The Intrapix test bench (Figure *7* and 8) is dedicated to the characterization of the Intra-Pixel Sensitivity Variations of astronomical detectors in a spectral range from visible to Infrared (12 µm). The measurement technique is based on the use of a grating, the Continuously Self Imaging Grating (CSIG) which generates N orders of diffraction. When it is illuminated with a plane wave, the transmittance produces a field whose intensity profile is a propagation-invariant biperiodic array of bright spots. The Fourier transform of the produced intensity pattern is composed of discrete spatial frequencies named CSIG harmonics. Those frequencies excited by the CSIG constitute the sparse data from which we can extract the Modulation Transfer Function parameter (MTF). The Pixel Response Function is deduced through the MTF[9].

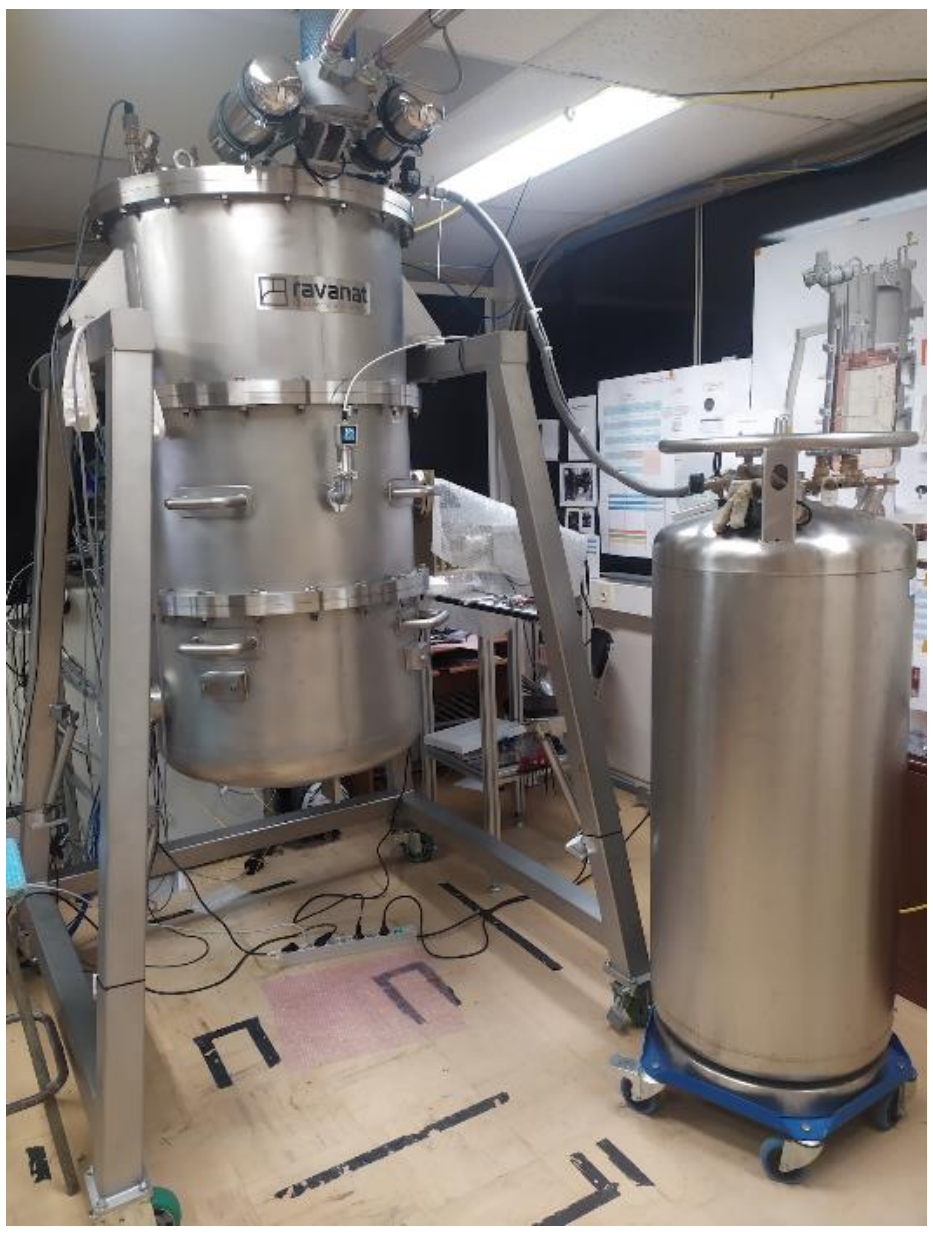

Figure 7: Intrapix platform at CEA Saclay

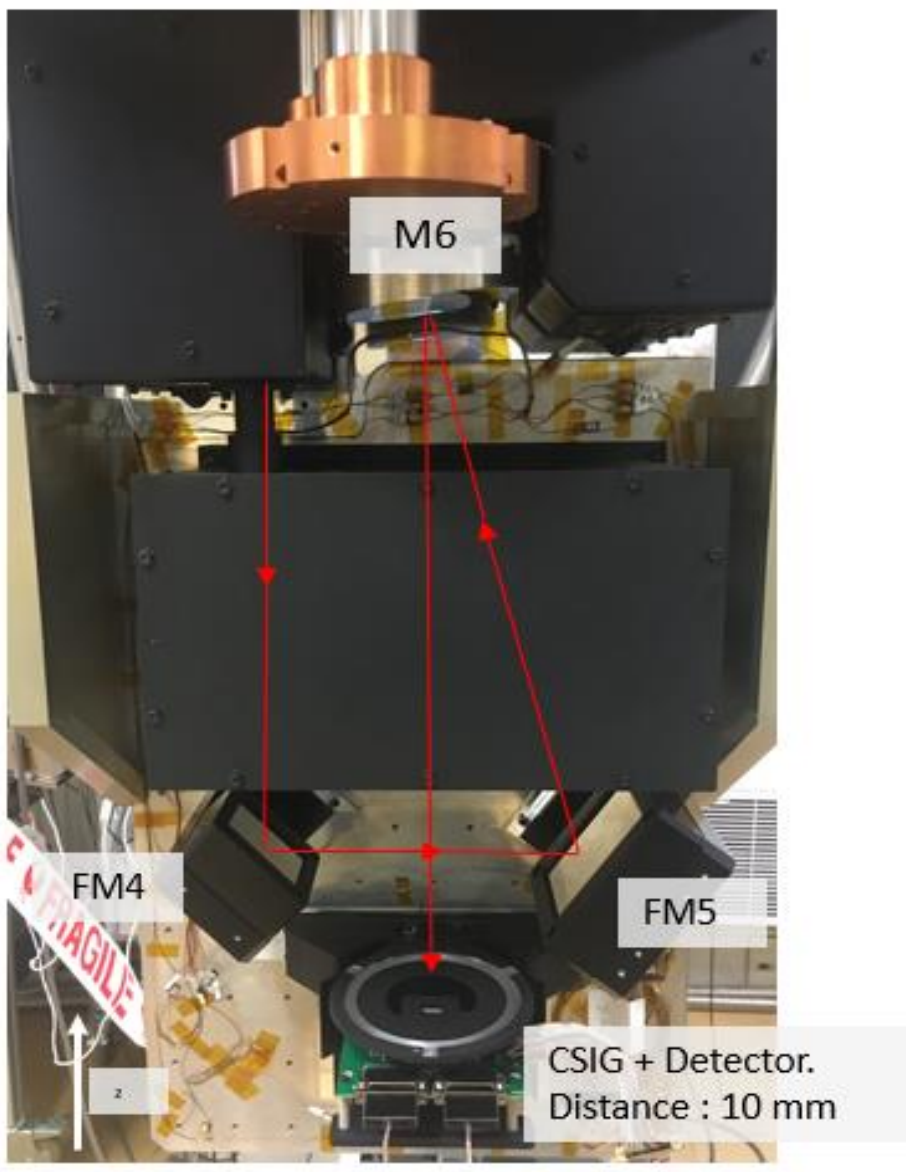

Figure 8: Inside Intrapix

### 3.2 Phase 2: Development of "Astro" packaging tailored for space-borne focal planes

Phase 2 aims to mitigate the main technical risks related to the space packaging and environmental qualification of large CMOS arrays for focal plane integration.

Thanks to a CEA heritage of space focal plane integration (HERSCHEL / PACS, EUCLID / VIS, JWST / MIRI, ARIEL / AIRS), a co-engineering approach between Pyxalis and CEA will be deployed. Design requirements will focus on compactness, ease of integration/removal, and compliance with space environment constraints, including mechanical vibrations, thermal variations, and radiation resistance. For astrometry focal plane, it is not useful to develop a buttable packaging. With more than 200 megapixels to bias and read out, the design (Figure *9*) provides easy access for connection and disconnection while keeping the bias and signal connectors separate. The electrical design and material selection will comply with ESA ECSS and NASA technical standards.

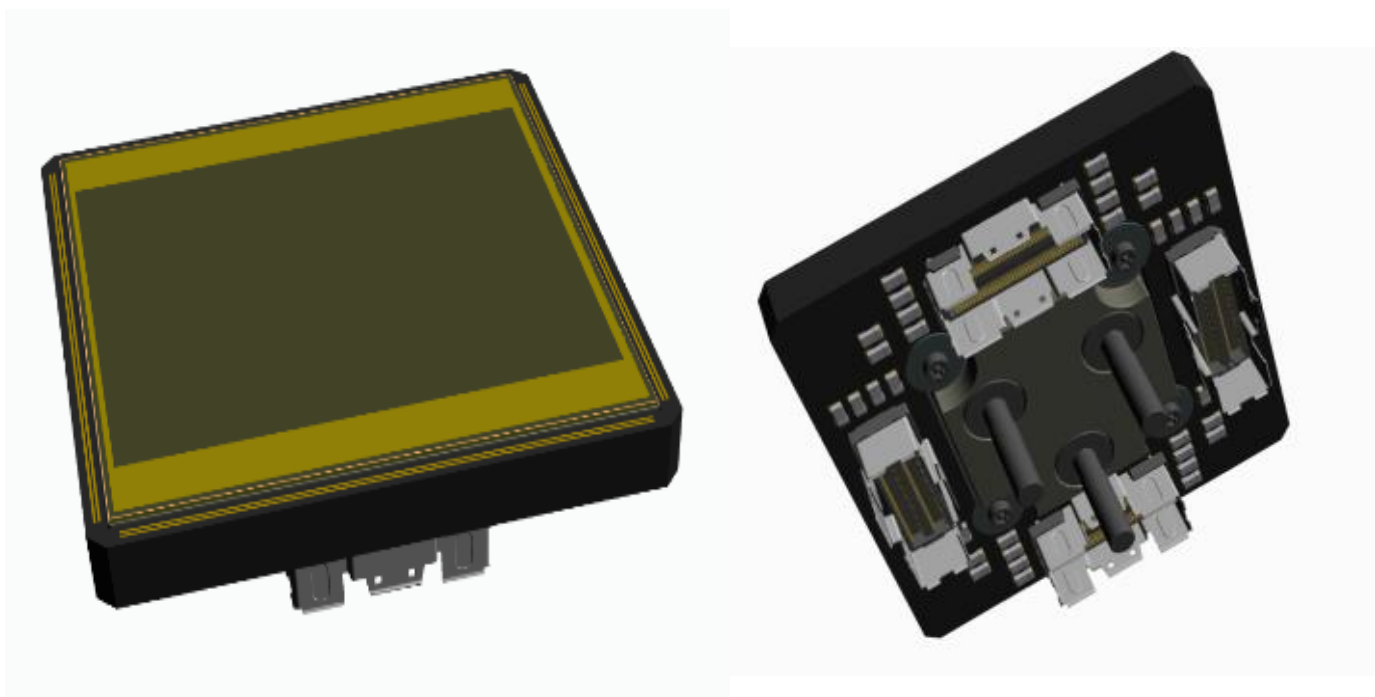

Figure 9: preliminary design of Astro GigaPyx 220 MP

Two detector array structural models identical in terms of thermal, mechanical and electrical interfaces will be delivered end of 2027 by the Pyxalis company. Following factory checks, an acceptance and test campaign will be conducted to verify compliance with requirements in four key areas:

(1) Mechanical metrology
    - Verification of external dimensions (82 mm x 80 mm x 50 mm)
    - Verification of mass (< 500 g)
    - Measurement of the first eigen frequency (> 1000 Hz)
(2) Thermal behavior
    - Measurement of the thermal gradient across the full sensitive area (<1K)
    - Verification of power dissipation (< 10 W)
(3) Electrical performance
    - Verification of connectivity through high density connectors suitable in space
    - Measurement of power supply characteristics and dynamic current profiles
(4) Integration and handling
    - Verification of how easy the mounting and dismounting is.
    - Verification of alignment features and interfaces

Successful completion will enable transition from “standard” to “Astro” model of a large CMOS sensor.

### 3.3 Phase 3: Detection chain performances validation

The objective of Phase 3 is to demonstrate that the performance of the 220 MP Gigapyx CMOS ‘Astro’ sensor is compatible with integration into a spaceborne focal plane and its ability to meet the requirements of a space mission. Based on the performances of the standard GigaPyx 220 MP CMOS sensor and the successful validation of the Astro packaging in Phase 2, two operational and functional detectors will be delivered second half of 2028. The ultimate goal is to demonstrate the detectors’ performance in a representative environment, with a focal plane assembly operated in a vacuum chamber in 2029.

Prior to the performance assessment of the complete focal-plane demonstrator—equipped with two 220 MP GigaPyx CMOS 'Astro' sensors and two structural models—the development and validation process will follow four key stages:

(1) Reception acceptance testing of each 220 MP GigaPyx “Astro” sensor
    - Mechanical inspection and dimensional verification of both sensors Thermal characterization, including power dissipation measurements and thermal cycling between 250 K – 320 K
    - Vibrations tests campaign to demonstrate stiffness above 1000 Hz.

(2) Radiation resistance tests campaign: standard levels will be considered to tests radiation resistance on a CMOS 220 MP Gigapyx “Astro” sample

(3) Validation of Astro packaging through detector performances characterization: the complete electro-optical characterization will be performed using the same test bench employed for the standard 220 MP GigaPyx CMOS sensor. The objective is to demonstrate that the Astro packaging preserves the intrinsic detector performance, including electro-optical characteristics, noise, linearity, and image quality.

(4) Development and performance validation at CEA of a dedicated detector readout chain (Figure *10*), leveraging CEA’s extensive expertise in space astrophysics instrumentation, with the electrical architecture and component selection defined to comply with applicable ECSS qualification requirements.

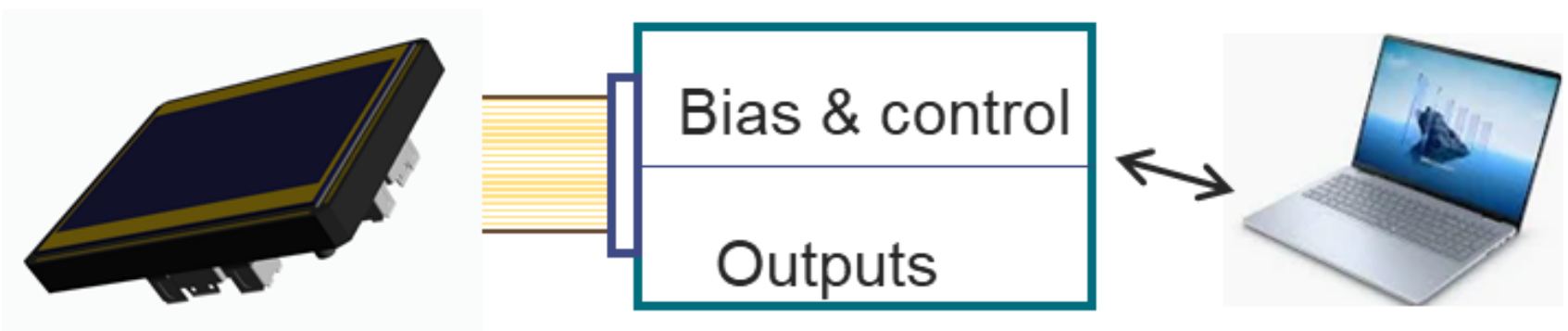

Figure 10 : CMOS 220 MP GigaPyx detection chain layout

### 3.4 validation CMOS 220 MP Gigapyx « Astro » in relevant environment

The objective of this phase is to validate that a partial focal plane equipped with the 220 MP GigaPyx Astro CMOS sensors, and integrated into an astrometric focal plane demonstrator, meets the specifications defined for the HWO-AGATE instrument. The system will be integrated with a calibration unit and operated inside a thermal-vacuum chamber to monitor and control the operating temperature.

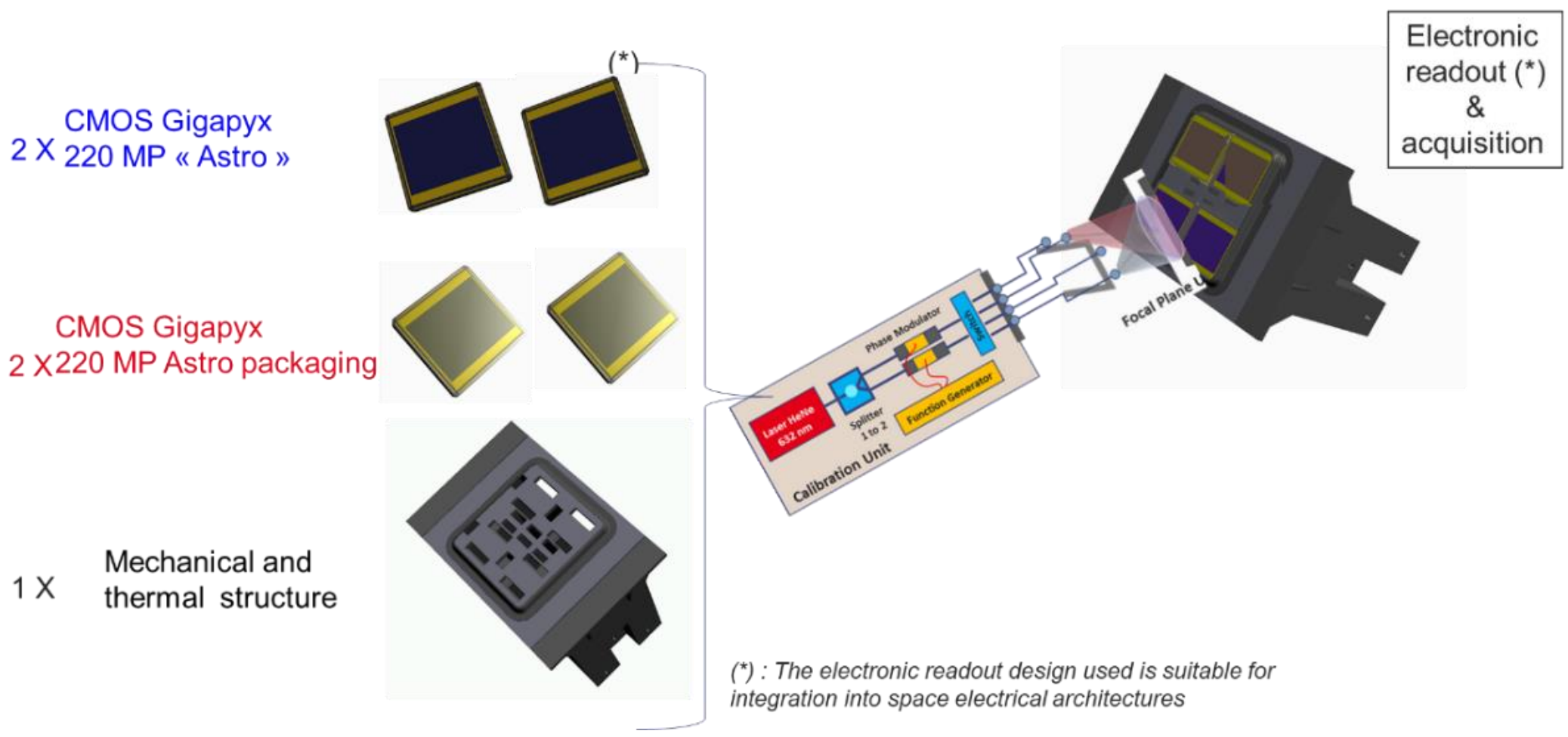

Figure 11: Astrometry instrument mock-up configuration for maturation phase

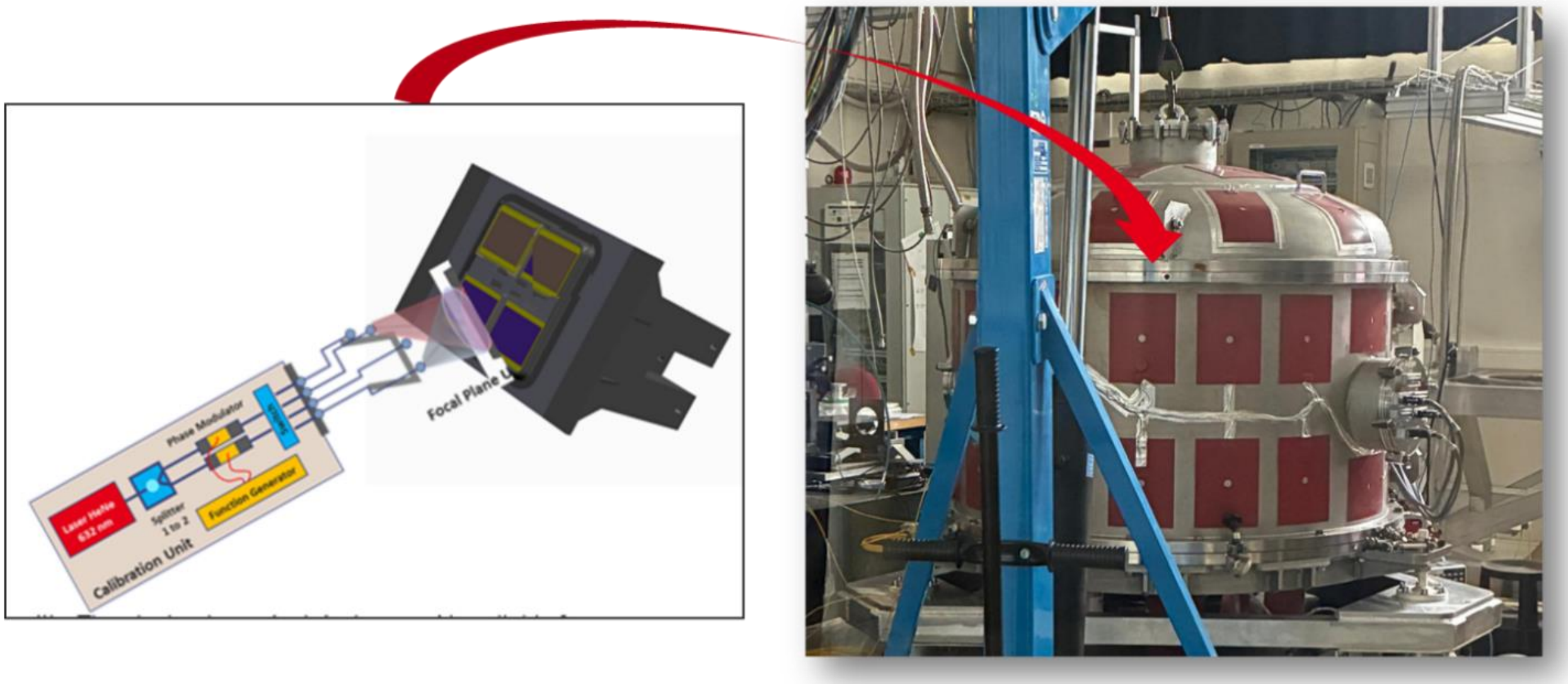

Figure 12 : Vacuum and thermal existing test facility

After the focal-plane integration, the GigaPyx Astro 220 MP sensor will be characterized as part of a partial focal plane array consisting of two active units and two structural models (see Figure *11*). All of these units will have identical electrical behavior and interfaces. The test campaign will evaluate performance under representative environmental conditions (thermal-vacuum). The test plan, which is currently under development, will provide the framework to qualify the calibration unit of the focal and the large-format CMOS sensors under operational conditions (Figure *12*).

## CONCLUSION

The proposed maturation program aims to demonstrate the feasibility of an integrated astrometric instrument architecture that combines large-format CMOS sensors, a representative focal-plane assembly, dedicated detector packaging, calibration capabilities, and a space-qualified detector readout chain. This architecture is designed to support high-speed data acquisition and onboard processing in accordance with applicable ECSS requirements.
Achieving this objective will significantly reduce the technological risks associated with future HWO astrometry instruments and pave the way for next-generation gigapixel-level focal plane instruments in space astronomy. The successful completion of this maturation plan will allow the 220 MP Gigapyx Astro large-format CMOS technology to reach NASA TRL 5 by the end of 2029, in time for the HWO Mission Conception Review.

## *ACKNOWLEDGEMENTS*

The authors gratefully acknowledge the contributions of the HWO-AGATE team of researchers and engineers for their invaluable support and collaboration throughout this work. This collaborative effort is supported by the CEA, CNRS, and Université Grenoble Alpes (UGA). The technology maturation program RETINA receives support from the French Government through the France 2030 investment plan, operated by the French National Research Agency under grant ANR-25-EXOR-0003. The work has been also partially supported by the LabEx FOCUS by the French National Research Agency through the grant ANR-11-LABX-0013 and in the framework of the "France 2030" program by the grant ANR-15-IDEX-02. The authors acknowledge financial support from the Centre Ntional d'Etudes Spatiales (CNES), France (ROR: https://ror.org/04h1h0y33), thanks to the Research and Technology (R&T) and mission preparation programs.